\documentclass[a4paper,fleqn,usenatbib]{mnras}
\usepackage[T1]{fontenc}
\usepackage{ae,aecompl}

\usepackage{graphicx}	
\usepackage{amssymb}	

\usepackage[usenames]{color}

\newcommand {\bc}{\begin{center}}
\newcommand {\ec}{\end{center}}
\newcommand {\be}{\begin{equation}}
\newcommand {\ee}{\end{equation}}
\newcommand {\beq}{\begin{eqnarray}}
\newcommand {\eeq}{\end{eqnarray}}

\newcommand {\ergs}{{\rm erg\ \rm s^{-1}}}

\newcommand {\comment}[1]{}

\renewcommand{\d}{{\rm d}}

\title[Optically thick accretion envelopes]
{Optically thick envelopes around ULXs powered by accreating neutron stars}
\author[A. A.~Mushtukov et al.] 
{Alexander~A.~Mushtukov,$^{1,2,3}$\thanks{E-mail: al.mushtukov@gmail.com (AAM)}  
Valery~F.~Suleimanov,$^{4}$
Sergey~S.~Tsygankov$^{5}$ 
\newauthor
and  Adam~Ingram$^{1}$\\ 
$^1$ Anton Pannekoek Institute, University of Amsterdam, Science Park 904, 1098 XH Amsterdam, The Netherlands \\
$^2$ Space Research Institute of the Russian Academy of Sciences, Profsoyuznaya Str. 84/32, Moscow
  117997, Russia\\
$^3$ Pulkovo Observatory, Russian Academy of Sciences, Saint Petersburg 196140, Russia \\
$^4$Institut f\"ur Astronomie und Astrophysik, Universit\"at T\"ubingen, 
    Sand 1, D-72076 T\"ubingen, Germany \\
$^5$Tuorla observatory, Department of Physics and Astronomy, University of Turku,
  V\"ais\"al\"antie 20, FI-21500 Piikki\"o, Finland} 
\date{Accepted ... Received 2016 December; in original form 2016 December .}

\pubyear{2016}

\begin{document}
\label{firstpage}
\pagerange{\pageref{firstpage}--\pageref{lastpage}}
\maketitle

\begin{abstract}
Magnetized neutron stars power at least some ultra-luminous X-ray sources. The accretion flow in these cases is interrupted at the magnetospheric radius and then reaches the surface  of a neutron star following magnetic field lines. Accreting matter moving along magnetic field lines forms the accretion envelope around the central object.
We show that, in case of high mass accretion rates $\gtrsim 10^{19}\,{\rm g\,s^{-1}}$ the envelope becomes closed and optically thick, which influences the dynamics of the accretion flow and the observational manifestation of the neutron star hidden behind the envelope. Particularly, the optically thick accretion envelope results in a multi-color black-body spectrum originating from the magnetospheric surface. The spectrum and photon energy flux vary with the viewing angle, which gives rise to pulsations characterized by high pulsed fraction and typically smooth pulse profiles. The reprocessing of radiation due to interaction with the envelope leads to the disappearance of cyclotron scattering features from the spectrum. We speculate that the super-orbital variability of ultra-luminous X-ray sources powered by accreting neutron stars can be attributed to precession of the neutron star due to interaction of magnetic dipole with the accretion disc.
\end{abstract}


\section{Introduction}
\label{intro}

Accretion onto a highly magnetized (surface magnetic field $B\gtrsim 10^{12}\,{\rm G}$) neutron star (NS) is governed by the magnetic field of the central object. Namely, the accretion disc is interrupted at the magnetosperic radius $R_{\rm m}$, where the magnetic pressure becomes comparable to ram pressure (due to Keplerian motion) of accreting material. Then the matter (a) follows magnetic field lines, (b) is accumulated at the magnetosphere \citep{1977PAZh....3..262S} or (c) is pushed away from the system depending on magnetic field strength, mass accretion rate and NS spin period \citep{2005A&A...431..597S,2016A&A...593A..16T}. If the matter continues its movement along magnetic field lines, it reaches the NS surface at small regions in the vicinity of the magnetic poles. There the matter releases its kinetic energy mostly in X-rays. This gives rise to X-ray pulsar (XRP)  phenomenon (see \citealt{2015A&ARv..23....2W} for recent review). The mass accretion rate, surface magnetic field structure and strength affect the geometry of the illuminating region \citep{1976MNRAS.175..395B,2015MNRAS.447.1847M}. Accretion luminosity of XRPs can be close or even higher than the Eddington luminosity. Recent discoveries of pulsating ultra-luminous X-ray sources (ULXs) show that XRP accretion luminosity can be as high as $10^{40} - 10^{41}\,\ergs$, which is a few hundreds times higher than the Eddington value for a NS \citep{2014Natur.514..202B,2016arXiv160907375I,2016arXiv160906538I,2016ApJ...831L..14F}.

ULXs powered by accreting NSs require special geometrical and physical conditions at the emitting region. It is known that a high mass accretion rate leads to the appearance of an accretion column above the surface of the NS \citep{1976MNRAS.175..395B,2015MNRAS.447.1847M}. The accretion flow is stopped at the top of the column in a radiation-dominated shock and then slowly settles down to the surface. The column can be as high as the NS radius \citep{2013ApJ...777..115P} and the material is confined there by magnetic field. It explains how accretion luminosity can principally exceed the Eddington limit \citep{2015MNRAS.454.2539M}. It is also important that the effective scattering cross-section and, therefore, radiation pressure are reduced by strong magnetic field \citep{2016PhRvD..93j5003M,2012PhRvD..85j3002M,2006RPPh...69.2631H}.

There are still debates on the magnetic field strength in recently discovered ULXs powered by magnetized NSs. 
The model of beamed X-ray sources fed at a super-Eddington accretion rate infers relatively weak magnetic field $\sim 10^{11}\,{\rm G}$ \citep{2016MNRAS.458L..10K}.
{Assuming torque equilibrium and solving the torque equation instead leads to $B$-field strength of $(2\div 7)\times 10^{13}\,{\rm G}$ \citep{2015MNRAS.448L..40E} or lower $\sim 10^{13}\,{\rm G}$ \citep{2015MNRAS.449.2144D} depending on model assumptions, which are essential near the equilibrium \citep{2016ApJ...822...33P}.} Transitions of ULX M82 X-2 between high and low states were interpreted as a ``propeller"-effect, requiring magnetar-like magnetic field $\sim 10^{14}\,{\rm G}$ \citep{2016MNRAS.457.1101T}. This is in good agreement with predictions of a model of accretion column \citep{2015MNRAS.454.2539M}. 

The dynamics of the flow in the accretion envelope (curtain) were assumed to be unaffected by radiation pressure so far. However, this is not the case for ULXs powered by accreting NSs. In this case the accretion flow between the inner disc radius and NS surface should be affected by radiation pressure. Moreover, simple estimations show that the accretion curtain at accretion luminosity $L\gtrsim 10^{40}\,\ergs$ becomes optically thick and shields the central object completely. It should influence the observational manifestation of bright XRPs. The possibility of NS shielding of by optically thick medium was supposed earlier \citep{2015MNRAS.448L..40E} in order to explain null result of search of pulsations in archival XMM-Newton observations of several ULXs performed by \citealt{2015A&A...579A..22D}.

In this paper, we discuss properties of accretion curtain at the magnetosphere of ultraluminous XRPs. We pay attention to its observational manifestation in spectral and timing properties of ULXs.

\section{Basic ideas}

\subsection{Accretion flow at the magnetospheric surface}

\begin{figure*}
\centering 
\includegraphics[width=12.cm, angle =0]{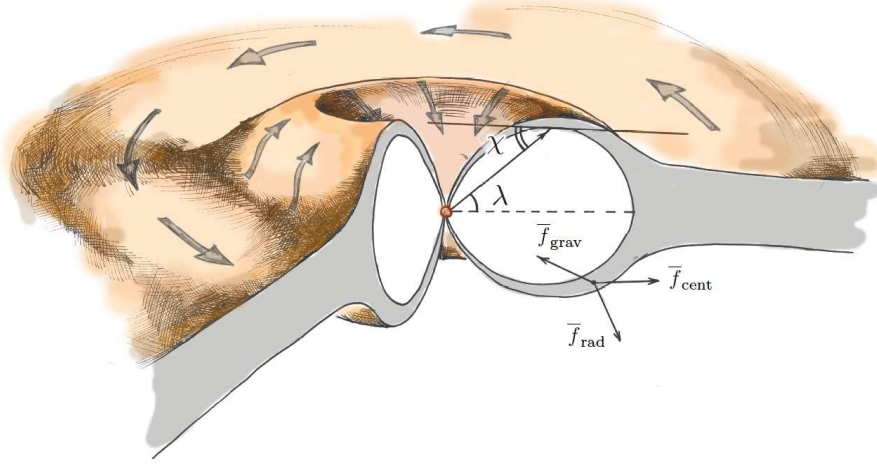} 
\caption{The structure of accretion flow in ULX powered by accreting NS. The geometrically thick accretion disc is interrupted at the magnetospheric radius and the accretion flow forms optically thick envelope. The inner part of dipolar magnetosphere is filled with hot thermolised photons. The observer detects the photons originated from the outer surface of magnetosphere.}
\label{pic:000}
\end{figure*}

The accretion disc is interrupted at the magnetospheric radius \citep{2014EPJWC..6401001L}, given by 
\be\label{eq:Rm}
R_{\rm m}=7\times 10^{7}\Lambda\, m^{1/7}R^{10/7}_{6}B^{4/7}_{12}L^{-2/7}_{39}\,\, \mbox{cm}, 
\ee
where $m=M/M_\odot$ is NS mass in units of solar masses, $B_{12}=B/10^{12}\,{\rm G}$ is surface magnetic field strength, $L_{39}=L/10^{39}\,\ergs$ is the accretion luminosity, $R_6=R/10^6\,{\rm cm}$ is the NS radius and $\Lambda$ is a constant which depends on the accretion flow geometry with $\Lambda=0.5$ being a commonly used value for the case of disc accretion.
From the magnetospheric radius the matter follows dipole magnetic field lines \citep{2002apa..book.....F}, which are described by the relation
\be
r(\lambda)=R_{\rm m}\cos^2 \lambda, 
\ee 
where $\lambda$ is the angular coordinate measured from the accretion disc plane and $r(\lambda)$ is a distance from the central object to a given point at the magnetic dipole line (see Fig.\,\ref{pic:000}).

The optical thickness of the flow at the magnetospheric surface is defined by the mass accretion rate $\dot{M}$, local velocity $v(\lambda)$ and mechanism of opacity. For the case of ionized gas at temperature $T>10^7 \,{\rm K}$ the opacity is defined by electron scattering (at lower temperature the Kramer opacity dominates with much higher cross-sections and optical thickness) and local optical thickness is
\be\label{eq:tau01}
\tau_{\rm e}(\lambda)\approx \frac{\kappa_{\rm e}\dot{M}d_0}{2S_{\rm D}v(\lambda)}\left(\frac{\cos\lambda_0}{\cos\lambda}\right)^3
\approx\frac{70\,L^{6/7}_{39}B^{2/7}_{12}}{\beta(\lambda)}\left(\frac{\cos\lambda_0}{\cos\lambda}\right)^3,
\ee
where $\beta(\lambda)=v(\lambda)/c$ is local dimensionless velocity along magnetic field lines, $S_{\rm D}\sim 10^{10}\,{\rm cm^2}$ is the area of accretion channel base at the NS  surface, $d_0$ is the geometrical thickness of the channel at the base, $\lambda_0\simeq \pi-(R/R_{\rm m})^{1/2}$ is the coordinate corresponding to the accretion channel base and $\kappa_{\rm e}=0.34\,{\rm cm^2\,\,g^{-1}}$ is the Thomson electron scattering opacity for solar abundances. 

The matter on the $B$-field lines moves under the influence of gravitational $\overline{f}_{\rm grav}$ and radiation $\overline{f}_{\rm rad}$ forces. 
If the photon flux is directed from the central object, the directions of the forces are opposite to each other. If  the local magnetic field strength is sufficiently high and is not significantly disturbed by radiative force, the dynamics of the material at the field lines is defined by the longitudinal component of the resulting force. The angle between the radius-vector pointed to a given point at the dipole field line and the tangent to the field line (see Fig.\,\ref{pic:000}) is
$$
\chi = \arctan(0.5/\tan(\lambda)).
$$
In the case of a rotating NS there is also a centrifugal force $\overline{f}_{\rm cent}$ in the corotating reference frame, which affects the dynamics of matter. The centrifugal force is directed perpendicular to the spin axis of a NS. Hence, the angle between the tangent to the field line and the centrifugal force is 
$$\xi=\pi+\chi-\lambda.$$

The resulting force gives an acceleration along magnetic field lines $a_{||}(\lambda)$ and the velocity of accretion flow is described by
\be\label{eq:beta}
\beta\frac{\d \beta}{\d\lambda}= \frac{R_{\rm m}\cos\lambda (4-3\cos^2\lambda)^{1/2}}{c^2}a_{||}(\lambda),
\ee
where $a_{||}(\lambda)$ is acceleration along the magnetic field lines caused by the resulting force.
This differential equation can be solved for known acceleration along magnetic field lines, given an initial velocity of material, mass accretion rate, surface magnetic field strength and spin period of the NS.

\subsection{Closed optically thick magnetospheric envelopes}
\label{sec:ClosedMag}
 
In the case of extremely high mass accretion rate
$\dot{M}\gtrsim 2.5\times 10^{18}\Lambda^{21/22}B_{12}^{6/11}m^{-5/11}$, which corresponds to accretion luminosity 
\be\label{eq:L_Azone}
L\gtrsim 3.4\times 10^{38}\,\Lambda^{21/22}B_{12}^{6/11}m^{6/11}\,\,\ergs, 
\ee
the accretion disc becomes radiation dominated and geometrically thick \citep{1973A&A....24..337S,2007ARep...51..549S}. 
In this case the accretion curtain is closed and radiation from the central object cannot leave freely the magnetosphere. Because the velocity along $B$-field lines cannot exceed the free-fall velocity, the optical thickness of the accretion curtain due to electron scattering $\tau_{\rm e}(\lambda)>\tau_{\rm e}(0)=L_{39}^{8/7}B_{12}^{-2/7}$ according to (\ref{eq:tau01}). Hence the accretion curtain is optically thick if
$L_{39}\gtrsim B_{12}^{1/4}$.
 At luminosity 
\be\label{eq:Lmax}
L\gtrsim 5.7\times 10^{39}\,\Lambda^{7/9}B_{12}^{4/9}m^{8/9}R_6^{1/3}\,\,\ergs
\ee
accretion disc thickness at the magnetospheric radius $R_{\rm m}$ is comparable to the size of NS magnetosphere \citep{1982SvA....26...54L}.
 
The photons from the central object are more likely reprocessed and reflected back into the cavity formed by the dipole magnetosphere than to immediately penetrate through the envelope. Each photon undergoes a number of scattering before it leaves the system. Hence, the radiation field is in equilibrium inside the magnetosphere and can be roughly described by black-body radiation of certain temperature $T_{\rm in}$, which is determined by the total accretion luminosity, the size of magnetosphere and its optical thickness. 

Let us consider a simplified problem of a point source of luminosity $L_{\rm ps}$ surrounded by spherical envelope of radius $R_{\rm sp}$ and optical thickness $\tau_{\rm sp}\gg 1$. Then the outer temperature of the spherical envelope is related to the inner temperature $T_{\rm in}$ as $T_{\rm out}\approx T_{\rm in}\tau_{\rm sp}^{-1/4}$. Because the total luminosity is conserved, the outer temperature is related to the luminosity and geometrical size of the envelope: $\sigma_{\rm SB}T^4_{\rm out}=L_{\rm ps}/(4\pi R^2_{\rm sp})$. Then the internal temperature is $T_{\rm in}=[\tau_{\rm sp}L_{\rm ps}/(4\pi\sigma_{\rm SB} R^2_{\rm sp})]^{1/4}$. In the case of a central source surrounded by a dipole surface, the internal temperature depends on the distribution of optical thickness over the surface.

The radiation force due to black-body radiation in the magnetospheric cavity  is directed perpendicularly to the magnetic field lines. As a result, the dynamics of the accretion flow along $B$-field lines is determined by gravity and centrifugal force only and described by the equation 
\beq
\label{eq:dbdl_2}
\beta\frac{\d \beta}{\d\lambda}&=&  \frac{GM(4-3\cos^2\lambda)^{1/2}}{c^2 R_{\rm m}\cos^3\lambda} \\
&& \times \left[\cos\chi + \left(\frac{\omega}{\omega^*_{\rm K}}\right)^2 \cos^7\lambda\cos\xi\right]. \nonumber
\eeq

Despite high optical thickness of the accretion curtain the photons penetrate through it after a number of scatterings. The time of photon diffusion through the accretion curtain is defined by local optical and geometrical thickness:
\beq\label{eq:t_diff}
t_{\rm diff}&\approx&\frac{\tau(\lambda)d(\lambda)}{c} \nonumber \\
&\sim & 2\times 10^{-9}\frac{L_{39}^{6/7}B_{12}^{2/7}d(\lambda)}{\beta(\lambda)}\left(\frac{\cos\lambda_0}{\cos\lambda}\right)^3 \,\,{\rm sec}.
\eeq 
If the photon diffusion time is much smaller than the dynamical time scale, the local flux at the outer side of a curtain is defined by the internal temperature and local optical thickness \citep{1969rtsc.book.....I}
\be
T_{\rm out}(\lambda)\simeq T_{\rm in}\tau^{-1/4}(\lambda). 
\ee
Because the temperature at the outer side of the curtain varies depending on the optical thickness,  the whole accretion curtain radiates multi-color black-body radiation.
The typical outer temperature can be roughly estimated as 
\beq\label{eq:Tout}
T_{\rm out}&\sim & \left(\frac{L}{\sigma_{\rm SB}4\pi R^2_{\rm m}}\right)^{1/4} \nonumber \\
&\approx & 0.5\,L_{39}^{11/28}B_{12}^{-2/7}m^{-1/14}R_6^{-5/7}\,\,\,{\rm keV}.	
\eeq
The internal temperature is higher and has to be found numerically. We see that the temperature in the envelope is $\gtrsim 1\,{\rm keV}$ for typical ULX luminosity and the opacity is, indeed, dominated by electron scattering.

The observed spectrum depends on the viewing angle as well because the observer sees different parts of the accretion curtain from different directions. 
The photon energy flux distribution over photon energy $E$ is defined by the temperature distribution over the visible parts of the envelope and given by 
\beq 
F_E=\frac{2E^3}{h^3 c^2}\frac{R_{\rm m}^2}{D^2}\int\d\varphi \int\d\lambda \cos i(\varphi,\theta)
\frac{\cos^4\lambda (4-3\cos^2\lambda)^{1/2}}{\exp\left[\frac{E}{kT_{\rm out}(\lambda)}\right]-1},\nonumber
\eeq
where the integrals are taken over the visible part of magnetic dipole surface, $i$ denotes the angle between the line of sight and local normal to the dipole surface and $D$ is the distance from the NS to observer. If the typical time of photon diffusion through the accretion curtain (\ref{eq:t_diff}) is comparable to the dynamical time-scale (which can be the case at high luminosity $\gtrsim few\times 10^{40}\,\ergs$), the final spectrum is disturbed.

The total luminosity of the accretion curtain can be obtained by integration of the flux over the dipole surface:
\beq
L&=&\frac{4\pi}{h^3 c^2}R^2_{\rm m} \int\limits_{0}^{\infty}\d E E^3\int\limits_{\lambda_1}^{\lambda_2}\d \lambda 
\frac{\cos^4\lambda (4-3\cos^2\lambda)^{1/2}}{\exp\left[\frac{E}{kT_{\rm out}(\lambda)}\right]-1}\nonumber  \\
&\simeq &6.3\times 10^{39}R^2_{\rm m,8}\nonumber \\
&&\times \int\limits_{0}^{\infty}\d E_{\rm keV} E_{\rm keV}^3\int\limits_{\lambda_1}^{\lambda_2}\d \lambda 
\frac{\cos^4\lambda (4-3\cos^2\lambda)^{1/2}}{\exp\left[\frac{E}{kT_{\rm out}(\lambda)}\right]-1} \nonumber 
\eeq
where $R_{\rm m,8}=R_{\rm m}/10^{8}\,{\rm cm}$ and $E_{\rm keV}=E/1\,{\rm keV}$.

It has to be noted that accretion from the disc likely results in an accretion channel, where the matter is confined to a narrow wall of magnetic funnel. In this case part of accretion column radiation can be radiated into the outer part of the magnetospheric cavity. This part of radiation is not reprocessed by the optically thick accretion curtain and is represented by a power-law spectrum.

The accretion disc by itself can contribute to the X-ray flux at extreme mass accretion rates. The effective temperature of the disc at the magnetospheric radius $R_{\rm m}$ (\ref{eq:Rm}) can be roughly estimated as 
\be
T_{\rm disc}\simeq 0.25\,L_{39}^{13/28}B_{12}^{-3/7}m^{-3/28}R_6^{-23/28} \,\,{\rm keV},
\ee
which is close to the temperature expected from the accretion envelope (\ref{eq:Tout}).
The energy spectrum of the accretion disc is likely given by milti-color black-body \citep{1973A&A....24..337S}, but can be disturbed due to advection.
As a result, one can expect an additional non-pulsating component component in spectrum of ULXs powered by accreting NSs.

The magnetic field strength at the magnetospheric surface should be high enough to confine the accretion flow affected by internal radiation pressure $P_{\rm rad}\approx a T_{\rm in}^4/3$. Then the magnetic pressure 
$$P_{\rm mag}(\lambda)=\frac{B^2(\lambda)}{8\pi}$$
should exceed $P_{\rm rad}$: $P_{\rm rad}/P_{\rm mag}\lesssim 1$.
The magnetic field pressure can be estimated from below by
$$P_{\rm mag}(\lambda =0)\simeq 3.4\times 10^{11} \Lambda^{-6}B_{12}^{-10/7}L_{39}^{12/7}m^{-6/7}R_{6}^{-18/7}.$$
Using estimation of the outer temperature (\ref{eq:Tout}) we get
\be
\frac{P_{\rm rad}}{P_{\rm mag}} \lesssim 7\times 10^{-2}\tau L_{39}^{-1/7}B_{12}^{2/7}m^{2/7}R_6^{-2/7},
\ee
which turns to $P_{\rm rad}/P_{\rm mag}\lesssim 0.07\,L_{39}m^{2/7}R_6^{-2/7}$ in the case of accretion flow of free-fall velocity. Therefore, the accretion envelope can be unstable at accretion luminosity above $10^{40}\,\ergs$, but this is beyond the scope of the present work.

\subsection{Interaction of the accretion curtain with X-ray photons in normal XRPs}

XRPs usually show hard energy spectra well fitted by a broken power law and complicated pulse profiles (see \citealt{2015A&ARv..23....2W} for recent review). Accreting NSs in normal XRPs are surrounded by an accretion curtain of intermediate optical thickness and relatively low temperature. The principal mechanisms of interaction between photons and accretion curtain are photoionization of heavy elements and Compton scattering with the energy recoil effect when the photon energy $E\gg k T_{\rm e}$  \citep{1976SvAL....2..111S}. The photons originating from the emitting region (hot spots or accretion column) are reprocessed partially. The photons experience $\sim \tau_{\rm e}^{2}$ scatterings on the average, where $\tau_{\rm e}$ is the optical thickness \citep{1969rtsc.book.....I,1972SvA....16...45I}. The initial energy of photons will be reduced due to the recoil effect from $E\gg k T_{\rm e}$ to $\sim \max(3k_{\rm B}T_{\rm e},m_{\rm e}c^2/\tau_{\rm e}^2)$, which leads to effective energy absorption by the accretion curtain \citep{1976SvAL....2..111S}. The absorbed energy is re-emitted in the soft X-ray energy band. Because the plasma layer occupies only some part of the magnetospheric surface, it affects X-ray pulsations and can cause the complicated structure of pulse profiles \citep{1976SvA....20..537B}. A detailed discussion of effects arising from accretion curtain in normal XRPs is beyond the scope of this paper.

\subsection{Power spectra affected by closed envelope}

XRPs commonly show fast aperiodic variability of X-ray flux over a broad frequency range (similar variability is shown by accreting black holes, see \citealt{2016AN....337..385I}). The variability is explained by the propagating fluctuations model \citep{1997MNRAS.292..679L}. The initial mass accretion rate variability is produced all over accretion disc due to MHD processes \citep{1991ApJ...376..214B} and results in mass accretion rate variability at the magnetospheric radius \citep{1997MNRAS.292..679L}. In the case of normal XRPs the observer detects photons directly from the emitting regions near the NS surface and variability of the mass accretion rate is directly imprinted in variability of observed X-ray flux \citep{2009A&A...507.1211R}, which provides the possibility of diagnostics of accretion disc. In case of extremely high mass accretion rate typical for ULXs the optically thick envelope screens the direct X-ray flux originating from the vicinity of the NS and reprocesses it. Observed variability is defined by variability of the mass accretion rate at the NS surface and variability of the local mass accretion rate and optical thickness at the magnetospheric envelope. As a result, aperiodic variability of ULXs is expected to be suppressed on time scales shorter than the timescale of matter travel from the disc to the NS surface.

\section{Model set up}

For simplicity and in order to get qualitative results we consider the accretion disc to be aligned with the equatorial plane of magnetic dipole. Inclining the magnetic dipole field changes the equations, but the main physical ideas remain the same.

We consider a Keplerian accretion disc, where the angular velocity is given by
\be
\Omega_{\rm K}=\left(\frac{GM}{R^3}\right)^{1/2}=11.5\,\left(\frac{m}{R^3_8}\right)^{1/2}\,\,{\rm rad\, s^{-1}}. 
\ee
The angular velocity of the magnetosphere is $\Omega=2\pi/ P$. If the mass accretion rate and accretion luminosity are higher than the limiting values corresponding to ``propeller"-effect (see, e.g., \citealt{2016A&A...593A..16T}):
\be
L_{\rm lim}\simeq 5\times 10^{37}\Lambda ^{7/2}B_{12}^2 P^{-7/3}m^{-2/7}R_6^5 \,\,\,\ergs,
\ee 
then the magnetospheric radius is smaller than the corotational radius and Keplerian angular velocity is higher than the angular velocity of NS magnetosphere. The accreting material looses its kinetic energy due to interaction with the magnetosphere. The lost energy is going into heat. The released energy is defined by the difference between Keplerian and magnetospheric angular velocities: $W_{\rm th}\propto (\Omega_{\rm K}-\Omega)^2$. The temperature caused by dissipation of kinetic energy is 
\be
T=\frac{1}{2}(\gamma - 1)T_{\rm vir}\left[1-\frac{\Omega}{\Omega_{\rm K}}\right]^2, 
\ee
where $\gamma$ is the adiabatic index and the virial temperature is defined as 
$$T_{\rm vir}=\frac{GM\mu}{R \mathcal{R}},$$
where $\mathcal{R}$ is the gas constant and $\mu$ the mean atomic weight per particle \citep{2010arXiv1005.5279S}.
Hence, the temperature can be estimated as 
$\min(T, \,T_{\rm disc}),$
where 
\beq
T\approx \frac{\gamma-1}{1-X}\Lambda^{-1}m^{6/7}L_{39}^{2/7}B_{12}^{-4/7}R_6^{-10/7}
\left[1-\frac{\Omega}{\Omega_{\rm K}}\right]^2\,\,{\rm keV} \nonumber,
\eeq
where $X$ is the hydrogen mass fraction.
The typical dimensionless thermal velocity of protons $\beta_{\rm th}=v_{\rm th}/c$ at the inner disc radii is given by
\beq\label{eq:beta_th}
\beta_{\rm th}&=&\frac{1}{c}\left(\frac{3k_{\rm B}T}{m_{\rm p}}\right)^{1/2} \nonumber \\
&\approx & 0.056\,\left(\frac{\gamma-1}{1+X}\right)^{1/2}\Lambda^{-1/2}m^{3/7} \nonumber \\
&& \times L^{1/7}_{39}B_{12}^{-2/7}R_{6}^{-5/7} \left| 1-\frac{\Omega}{\Omega_{\rm K}} \right|, 
\eeq
We take this velocity as the initial velocity of the accretion flow along the magnetic field lines.

The accretion disc has a certain geometrical thickness, which is defined by hydrostatic balance in the vertical direction \citep{1973A&A....24..337S} and affected by the origin of opacity \citep{2007ARep...51..549S}. The disc can be divided into three zones according to the dominating pressure and opacity sources. Gas pressure and  Kramer opacity dominate in the outer regions of the accretion disc (C-zone). The relevant disc scaleheight is given by
\be
\left(\frac{H}{r}\right)_{\rm C}= 0.056\,\alpha^{-1/10}L_{39}^{3/20}m^{-21/40}R_6^{3/20}r_8^{1/8},
\ee
where $r_{8}=r/10^{8}\,{\rm cm}$ is the dimensionless radial coordinate. 
Gas pressure and electron scattering dominate in the intermediate zone (B-zone), where
\be
\left(\frac{H}{r}\right)_{\rm B}= 0064\,\alpha^{-1/10}L_{39}^{1/5}m^{-11/20}R_6^{1/5}r_8^{1/20},
\ee
and radiation pressure may dominate in the inner zone (A-zone) where
\be\label{eq:Azone}
\left(\frac{H}{r}\right)_{\rm A}= 0.1 \frac{L_{39}R_{6}}{m}r_{8}^{-1}.
\ee
The boundary between B- and C-zones is at $r_{\rm BC}\approx 9\times 10^8\, L_{39}^{2/3}m^{-1/3}R_{6}^{2/3}\,{\rm cm}$, while the boundary between A- and B-zones is at $r_{\rm AB}\approx 2.2\times 10^8\,L_{39}^{16/21}m^{-3/7}R_{6}^{16/21}\,{\rm cm}$.

At mass accretion rates above $few\times 10^{18}\,\,{\rm g\,s^{-1}}$ (see equation (\ref{eq:L_Azone})) the disc is interrupted at the radiation dominated A-zone. According to (\ref{eq:Azone}) geometrical thickness can be comparable to the magnetospheric radius at $L\gtrsim 6\times 10^{39}\,\Lambda^{7/9}B_{12}^{4/9}m^{8/9}R_6^{1/3}$ \citep{2015MNRAS.454.2539M,1982SvA....26...54L}, but increase of accretion disc thickness can be stopped if accretion disc becomes advection-dominated \citep{2016A&A...587A..13L}. Hence, the geometrical thickness of accretion disc given by (\ref{eq:Azone}) can be overestimated.

The geometrical thickness gives the initial coordinate $\lambda$ from which the accretion flow starts its motion along magnetic field lines. We assume that the accretion flow starts with typical thermal velocity defined by interaction between the accretion disc and the magnetoshpere at $R_{\rm m}$ (\ref{eq:beta_th}). It might be important for the case of extremely high mass accretion rates, when the disc is interrupted at its radiation dominated zone (A-zone) and geometrically thick there.

\begin{figure}
\centering 
\includegraphics[width=7.7cm]{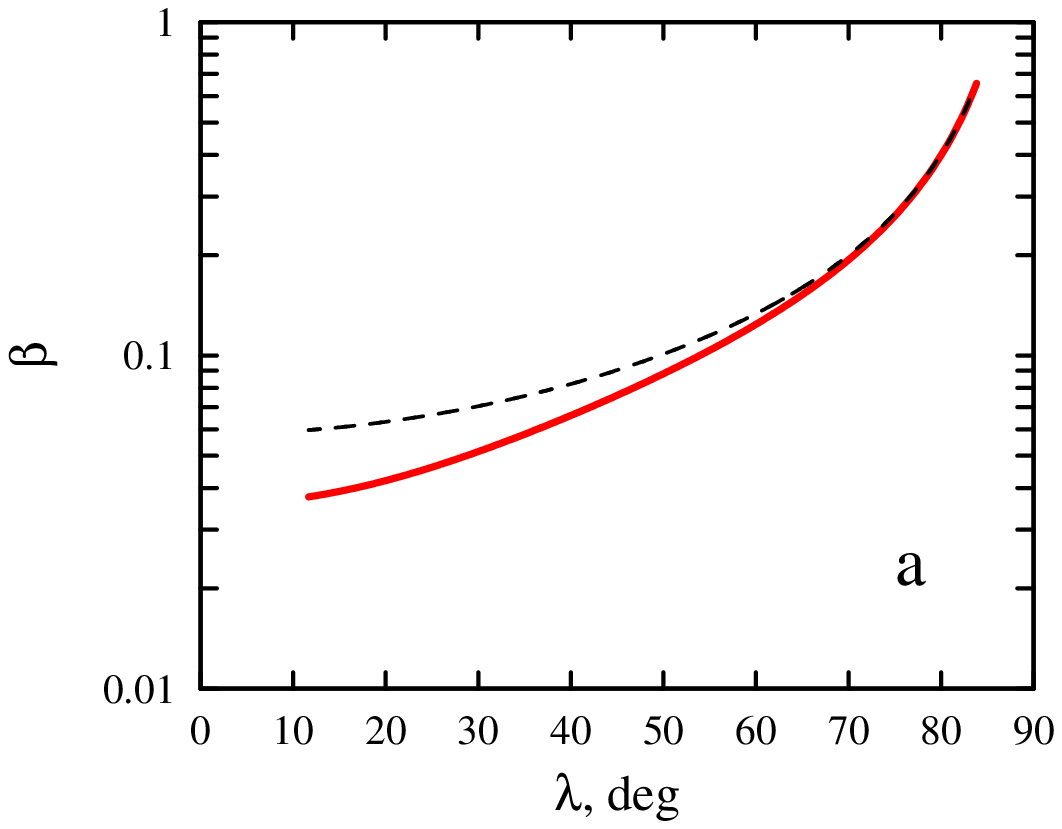} 
\includegraphics[width=7.7cm]{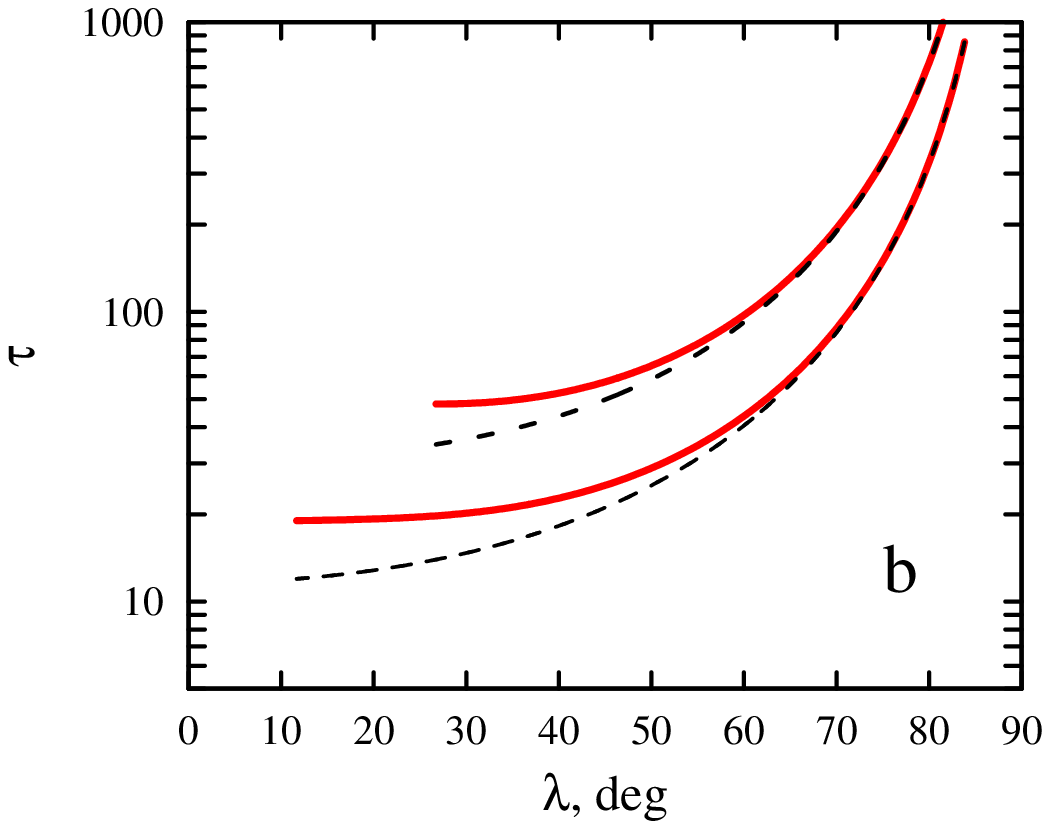} 
\includegraphics[width=7.7cm]{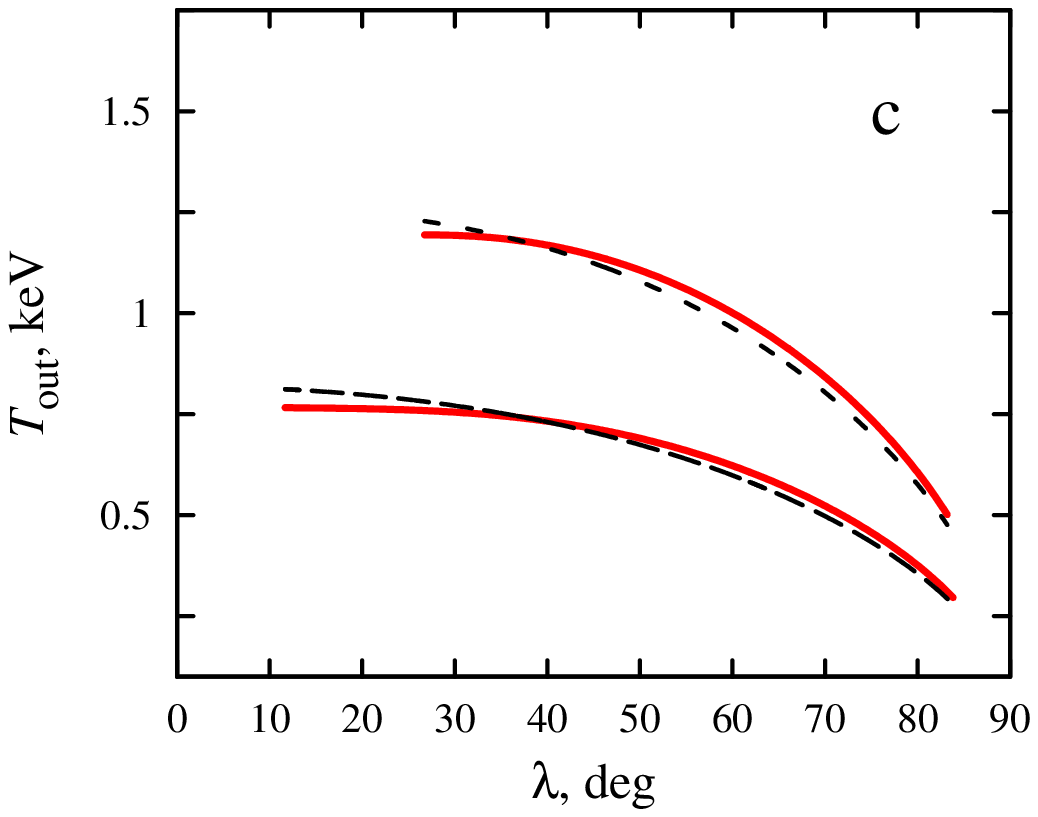} 
\caption{The dimensionless velocity (a), optical thickness (b) and effective outer temperature (c) at the dipole surface of the accreting NS as a function of coordinate $\lambda$. Red solid and dashed black lines are given for NS spinning with period $P=1\,{\rm sec}$ and non-rotating NS respectively. 
The accretion flow velocity is not affected significantly by the mass accretion rate and the curves at (a) correspond to accretion luminosity $5\times 10^{39}\,\ergs$. Two sets of curves on (b) and (c) plots are given for accretion luminosity of $5\times 10^{39}\,\ergs$ \textit{(lower)} and $10^{40}\,\ergs$ \textit{(upper)}. 
Magnetic field strength is $B=10^{13}\,{\rm G}$. The parameters used in the calculations: $M=1.4M_\odot$, $R=10^6\,{\rm cm}$, $\Lambda=0.5$.}
\label{pic:examp}
\end{figure}

\section{Numerical results}

We solve equation (\ref{eq:dbdl_2}) numerically for a given NS mass, spin period and surface magnetic field strength, taking the initial velocity to be equal to the  thermal velocity of protons at the inner radius of accretion disc (\ref{eq:beta_th}). The initial coordinate is defined by the geomerical thickness of accretion disc $\lambda_{\rm ini}={\rm atan}(0.5 H/R)$ at the magnetospheric radius.

The typical velocity profiles and corresponding optical thickness distributions over the magnetospheric surface are given by Fig.\,\ref{pic:examp}a,b.
The mass accretion rate and optical thickness define the surface temperature $T_{\rm out}$ (see Fig.\,\ref{pic:examp}c). The rotation of NS affects the surface temperature only slightly.
According to numerical solutions of equation (\ref{eq:dbdl_2}) the typical time scale of motion of material between the magnetospheric radius and NS surface is $\sim (0.1\div 1)\,{\rm sec}$. The accretion flow receives the major part of final kinetic energy in the very vicinity of a NS, where the influence of radiation and centrifugal forces is negligible. As a result, the final velocity of matter near NS surface is affected by mass accretion rate only slightly and earlier theoretical constructions on the structure of X-ray flux forming region \citep{1976MNRAS.175..395B,2015MNRAS.447.1847M,2015MNRAS.454.2539M,2015MNRAS.454.2714M} are not influenced by dynamics in the accretion curtain. 

\begin{figure}
\centering 
\includegraphics[width=9.cm]{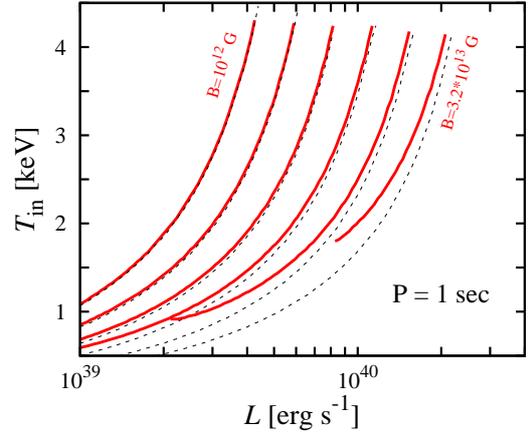} 
\caption{The internal temperature $T_{\rm in}$ as a function of accretion luminosity is shown by red solid lines for NS spinning with period $P=1\,{\rm sec}$ and by black dashed lines for non-rotating NS. Different curves correspond to different magnetic field strength: $10^{12},\,2\times 10^{12},\,4\times 10^{12},\,8\times 10^{12},\,1.6\times 10^{13},\,3.2\times 10^{13}\,{\rm G}$ (from left to right). The upper ends of red curves are defined by condition of super-critical disc at the magnetospheric radius (defined by the disc scale height becoming comparable to the magnetospheric radius). The parameters used in the calculations: $M=1.4M_\odot$, $R=10^6\,{\rm cm}$, $\Lambda=0.5$.}
\label{pic:T_in}
\end{figure}

The internal temperature $T_{\rm in}$ depends on the mass accretion luminosity and magnetic field strength (see Fig.\,\ref{pic:T_in}). The temperature is affected by the spin period of the NS because rotation influences the velocity (and therefore the local optical thickness) of accretion flow due to the centrifugal force.

\begin{figure}
\centering 
\includegraphics[width=9.cm]{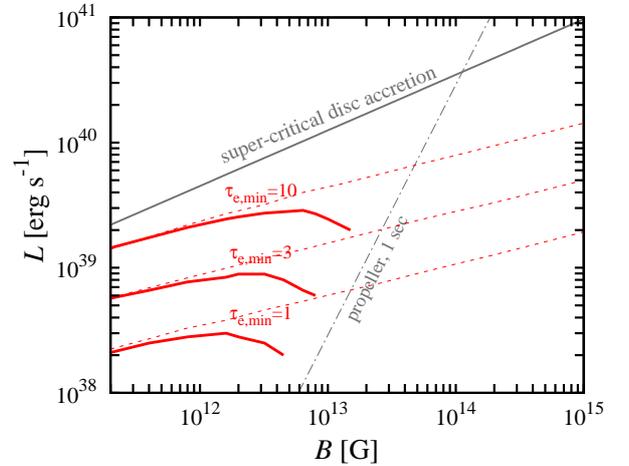} 
\caption{The accretion luminosities where the minimal optical thickness of the envelope due to electron scattering is equal 1, 3 and 10, are given by red solid lines (spin period of NS is taken to be $P=1\,{\rm sec}$) and red dashed lines (non-rotating NS). The grey dashed-dotted line shows the limiting luminosity due to the propeller effect for spin period $P=1\,{\rm sec}$. The centrifugal force becomes essential when the accreting NS is close to the limiting propeller luminosity: the force slows down the accretion flow and makes it optically thicker. The grey solid line limits the region of super-critical disc accretion given by (\ref{eq:Lmax}). The parameters used in the calculations: $M=1.4M_\odot$, $R=10^6\,{\rm cm}$, $\Lambda=0.5$.}
\label{pic:005}
\end{figure}

Because of the geometry of the accretion flow, the optical thickness can be high enough to cause advection of photons. The advection process becomes important when the time of photon diffusion (\ref{eq:t_diff}) becomes comparable to the time over which the material passes a distance equal to the local geometrical thickness of the envelope: $t^*(\lambda)=d(\lambda)/(c\beta(\lambda))$. It happens at $\cos\lambda \lesssim 4\cos\lambda_0 L_{39}^{2/7}B_{12}^{2/21}$. Therefore, the advective zone is located very close to the NS surface even at accretion luminosity $L\sim 10^{39}\div 10^{40}\,\ergs$. We do not take advection in the accretion envelope into account. 
 
The matter velocity of the accretion flow is significantly smaller than free-fall velocity all over the envelope except the regions close to NS surface. As a result, the estimation of optical thickness from the assumption of free-fall velocity is underestimated and the accretion curtain becomes optically thick already at $L\gtrsim 10^{39}\,\ergs$ (see Fig.\,\ref{pic:005}). Therefore, ULXs powered by accretion onto magnetized NS have to be surrounded by the optically thick envelope. The minimal optical thickness on the magnetosphere is affected significantly by centrifugal force if the accretion luminosity becomes comparable to the limiting propeller luminosity: the centrifugal force reduces the velocity along magnetic field lines, which leads to increasing of optical thickness.

\section{Summary}
\label{sec:Summery}

The accretion onto magnetized NS goes through the accretion curtain at the magnetosphere. In the case of extremely high mass accretion rates typical for ULXs, the accretion curtain becomes closed and optically thick (see Fig.\,\ref{pic:005}). This affects the dynamics of matter in the curtain and the basic observational manifestation of the accreting NS at luminosity above $few\times 10^{39}\,\ergs$. 

The closed accretion curtain intercepts and reprocesses the initially hard radiation from the central object into black-body like radiation of temperature which depends on the local optical thickness and total accretion luminosity (see Fig.\,\ref{pic:examp}c). Hence the accretion curtain manifests itself by multi-color black-body spectrum, which varies with the viewing angle and can be dominated by the curtain temperature. In this case the total spectrum can be fitted by double black-body, where the harder component corresponds to accretion envelope and low energy componenet corresponds to an advection dominated accretion disc.
Variations of viewing angle due to NS spin lead to variations of the observed spectrum and cause observed pulsations. 
Because the accretion disc provides a non-pulsating low energy component in the spectrum, the pulsed fraction is expected to be higher in the higher energy band. These statements are in agreement with observational data: the spectrum of pulsating ULX-1 in NGC 5907 is well-described by multi-color black-body \citep{2016arXiv161000258F}, while the pulsed fraction is, indeed, higher in the higher energy band. It is interesting, that the spectra of several ULXs are well fitted with a double black-body model \citep{2006MNRAS.368..397S,2016arXiv160903941K}.

At larger timescales the viewing angle can also vary due to precession of the magnetic dipole \citep{1980SvAL....6...14L} giving rise to super-orbital variability. The precession period of the NS can be roughly estimated:
\be
t_{\rm pr}\approx 1.5\times 10^4\,\frac{\mu_{30}^{-2}I_{45}R_{\rm m,8}^3 P^{-1}} {\cos\psi  (3\cos\zeta -1)}\,\,\,{\rm years},
\ee
where $\psi$ is the angle between the normal to the accretion disc plane and NS spin axis, $\zeta$ is the angle between the spin axis and magnetic field axis, $\mu_{30}=B_{12}R_6^3$ is NS magnetic moment and $I_{45}$ is NS moment of inertia. If NS has surface dipole magnetic field $>10^{14}\,{\rm G}$ the precession period can be about a year. {It is interesting that a few ULXs including bright pulsating ULX-1 in galaxy NGC 5907 \citep{2016arXiv160907375I} show modulation of a photon flux with typical period of a few months: pulsating M82 X-2 - 55 days \citep{2016MNRAS.461.4395K}, M82 X-1 - 62 days \citep{2006ApJ...646..174K}, pulsating ULX NGC 7793 P13 - 65 days and $3000$ days \citep{2017arXiv170102449H}, pulsating ULX-1 NGC 5907 - 78 days \citep{2016ApJ...827L..13W}, NGC 5408 X-1 - 115 days and 243 days \citep{2009ApJ...706L.210S,2012RAA....12.1597H}, HLX ESO 243-39 - 375 days \citep{2011ApJ...743....6S}. }

Finally we conclude that

(i) Optically thick accretion envelope affects the observed spectra of ULXs because of reprocessing of X-ray radiation from the central object. In the case of a closed accretion curtain, the spectrum is described by a multi-color black-body, which can vary with	viewing angle. It is likely that the multi-color black-body spectrum of the accretion envelope is dominated by the curtain temperature.

(ii) Optically thick envelope makes cyclotron lines originating from the central object undetectable.

(iii) The curtain affects the observed pulse profile. The observer see photons, that have been reprocessed by the envelope and re-emitted from a much more extended photosphere of a geometrical size comparable to the geometrical size of NS magnitosphere. This explains a difference between typical pulse profiles of XRPs and pulsating ULXs: in case of normal XRPs, observer detects photons from the central object and the pulse profile is disturbed by the accretion curtain, while in the case of ULXs the pulse profile is totally defined by emission from the envelope.

(iv) In the case of normal XRPs velocity of the matter in the accretion curtain is almost unaffected by radiation pressure. In the case of ULXs powered by accretion onto a NS, the velocity of material on the magnetosphere is different, but the velocity near the NS surface is still close to free-fall velocity. As a result, the developed theoretical models of the geometry of illuminating regions are valid \citep{1976MNRAS.175..395B,2015MNRAS.454.2539M}.

\section*{Acknowledgments}

This research was supported by  the Russian Science Foundation grant 14-12-01287 (AAM) and the German research Foundation (DFG) grant
WE 1312/48-1 (VFS). Partial support comes from the EU COST Action MP1304 ``NewCompStar".


{

}

\bsp	
\label{lastpage}
\end{document}